\documentclass[aps,prb,preprint,onecolumn,superscriptaddress]{revtex4-2}
\usepackage{amssymb,amsmath,amsfonts}
\usepackage{graphicx} 
\usepackage{float} 
\usepackage{xcolor}
\usepackage{soul}

\usepackage[
colorlinks=true,bookmarks=false,citecolor=blue,urlcolor=blue
]
{hyperref}

\usepackage{natbib}

\graphicspath{ {./Figures/} }

\renewcommand{\Re}{\frak{R}\mathrm{e} }
\renewcommand{\Im}{\frak{I}\mathrm{m} }
\newcommand{\DM}{\Delta_\mathrm{max} }

\newcommand{\eps}{\varepsilon}
\newcommand{\w}{\omega}

\begin{document}

\title{
The influence of shot noise on the performance of phase singularity-based refractometric sensors
}

\author{Valeria Maslova}
\affiliation{Moscow Center for Advanced Studies, Moscow, 123592, Russia}

\author{Georgy Ermolaev}
\affiliation{Emerging Technologies Research Center, XPANCEO, Internet City, Emmay Tower, Dubai, United Arab Emirates}

\author{Evgeny S. Andrianov}
\affiliation{285011 Dukhov Research Institute of Automatics (VNIIA) , 22 Sushchevskaya, Moscow 127055, Russia}
\affiliation{Moscow Center for Advanced Studies, Moscow, 123592, Russia}

\author{Aleksey V. Arsenin}
\affiliation{Emerging Technologies Research Center, XPANCEO, Internet City, Emmay Tower, Dubai, United Arab Emirates}

\author{Valentyn S. Volkov}
\affiliation{Emerging Technologies Research Center, XPANCEO, Internet City, Emmay Tower, Dubai, United Arab Emirates}

\author{Denis G. Baranov}
\email[]{baranov.mipt@gmail.com}
\affiliation{Moscow Center for Advanced Studies, Moscow, 123592, Russia}

\begin{abstract}
Topological singularities of optical response functions -- such as reflection amplitudes -- enable elegant practical applications ranging from analog signal processing to novel molecular sensing approaches.
A phase singularity-based refractometric sensor monitors the rapidly evolving \emph{argument} of the optical field near the point of phase singularity, in contrast to the reflection zero in traditional surface plasmon polariton sensors. This raises a natural question: What happens with the sensitivity and resolution of such a sensor when it operates close to a zero of the response function, where the detected signal may be greatly influenced by various noise sources?
In this paper, we systematically study the effect of the shot noise on the performance of a generic phase singularity-based refractometric sensor.
We develop a theoretical model of a spectroscopic ellipsometry-based system operating near a phase singularity and couple the macroscopic optical picture of the detection with a quantum shot noise model.
Within the developed model, we illustrate how the shot noise of the detector comes into play and study its effect on the sensitivity and resolution of the refractometric sensor.
Our results suggest that such an ellipsometry-based phase singularity sensor remains stable even in the presence of shot noise near the point of zero reflection.
\end{abstract}

\maketitle
\newpage

\section{Introduction}

The ability to detect trace amounts of various organic and inorganic substances, viruses, and microscopic living organisms is crucial for effective health monitoring \cite{borisov2008optical, turner2013biosensors}. Nanophotonics, the study of light-matter interaction at the nanometer scale, offers a range of techniques for such detection. These include surface-enhanced Raman scattering \cite{sharma2012sers, pilot2019review, langer2019present}, surface plasmon resonance \cite{anker2008biosensing, Kabashin2009, Mayer2011, Zhang2015, Lee2016, Sreekanth2016, jeong2016dispersion},  dielectric nanoparticle resonance \cite{bontempi2017highly, bosio2019plasmonic, krasnok2018spectroscopy}, high-Q lasing structures \cite{He2011}, and metasurface-assisted hyperspectral infrared imaging \cite{Tittl2018, leitis2019angle, Yesilkoy2019, Tseng2021}. Many of these methods are refractometric, measuring changes in the refractive index of the surrounding analyte medium induced by the substance, thereby allowing the evaluation of the substance's concentration. One should separately mention various approaches and challenges in chiral sensing, where changes in circular dichroism are monitored in order to evaluate the disbalance between right-handed and left-handed enantiomers of a chiral racemic \cite{Tang2011, Hendry2010, Govorov2010, Graf2019, kim2022enantioselective, warning2021nanophotonic, both2022nanophotonic}. 

Novel generalizations of the existing approaches also appear, such as nonlinear plasmonic sensing \cite{Mesch2016} or quantum sensing \cite{fan2015quantum, Lee2016}. Recently, more exotic approaches based on exceptional points of non-Hermitian systems \cite{Wiersig2014, Wiersig2016, Chen2017, Hodaei2017, Jing2018, xu2024single} and phase singularities of optical fields have been intensively explored \cite{Kravets2013, Malassis2014}. 
The latter relies on a linear optical system with a phase singularity of its response function, such as reflection -- the point in the parameter space of wavelength and incidence angle with zero amplitude \cite{sreekanth2018ge2sb2te5, Berkhout2019, Liu2023, thomas2022all, cusworth2023topological, maslova2024topological}. The argument of the signal varies rapidly in the vicinity of this point, thus allowing enhanced sensitivity \cite{vasic2014enhanced, sreekanth2018biosensing, ermolaev2022topological, tselikov2023topological, zhu2024label}.
The realization of phase singularities is feasible in simple planar structures \cite{ermolaev2022topological}, and even at a single anisotropic interface \cite{zhu2024label}, rendering the idea suitable for widespread use.


Despite the intuitive appeal of the idea, the use of a singularity come at the cost of low intensity of the signal close to the point of zero reflection.
The optical signal is subject to noises of various origins, including the thermal noise of the background \cite{Baranov2019}, the fluctuating density of the analyte medium \cite{antosiewicz2016multiscale}, or the shot noise due to the discrete nature of photons \cite{Lee2016, antosiewicz2016multiscale, MandelLeonardandWolf1995, chong2013noise}. Correspondingly, these noises may introduce significant uncertainties in the argument of the detected signal at the point of zero response:
\begin{equation*}
    \psi = \arctan \frac{\Im E + \delta_i}{\Re E + \delta_r},
    \label{Eq_1}
\end{equation*}
thus completely diminishing the sensitivity, which is a common challenge of the exceptional points-based sensors \cite{Lau2018, Langbein2018, Mortensen2018, Duggan2022}. As a result, it poses a fundamental question of the robustness of a phase singularity-based sensor against such noises.

In this paper, we address the issue of shot noise and its impact on the performance of a generic phase singularity-based refractometric sensor.
To this end, we develop a universal theoretical model of the phase singularity-based sensor that incorporates the shot noise arising at the detection stage, and takes into account the experimental details of commercial detection schemes. 
Using this model, we study the effect of finite exposure and the choice of the initial calibration curve on the sensitivity and resolution of the detector. Our results indicate that this type of optical sensing is robust against shot noises. Furthermore, our model has predictive capabilities for finding optimal phase singularity-based sensor parameters. Therefore, our findings provide an advanced tool for the development of current and next-generation biosensors.

\section{Results}

\subsection*{Basic principles of phase singularity-based sensors}

\begin{figure}[t!]
\centering\includegraphics[width=1.0\textwidth]{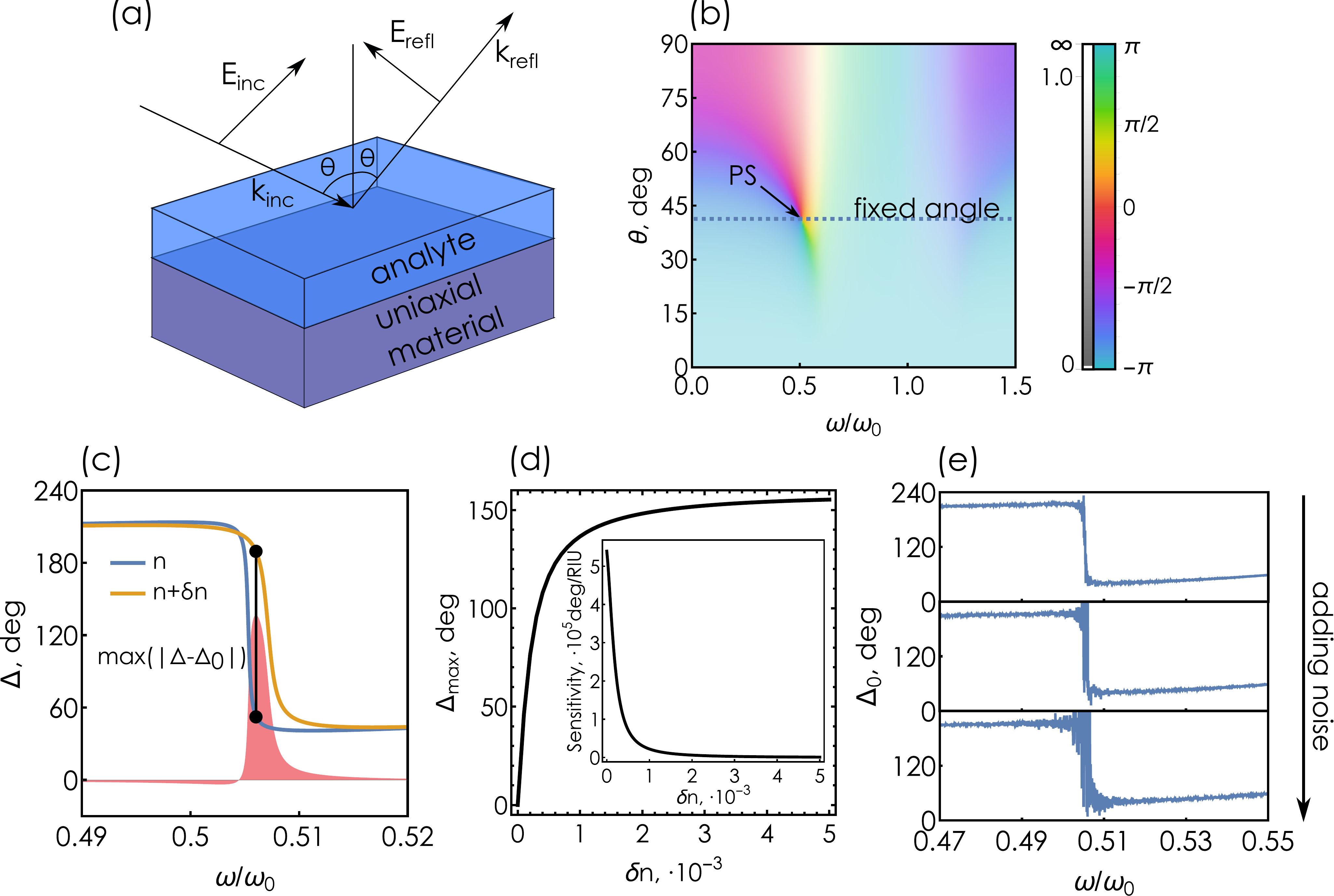}
\caption{\textbf{Basic picture of of phase singularity sensing.} 
(a) Sketch of the refractometric sensor under study: a uniaxial absorbing material covered with an optically thick layer of analyte. The system is illuminated with a combination of $s$- and $p$-polarized waves; the amplitude and the phase of the reflected wave is read out using spectroscopic ellipsometry.
(b) Plot of the complex-valued ellipsometric response function $\rho = r_p/r_s$ for the system in (a) calculated for water reference solution, $n=1.33$, and the uniaxial material is characterized by $\eps_{\infty} = 1$, $f = 0.5$, $\gamma = 0.1$, $\eps_{||}=1.5+0.1i$. 
(c) Frequency dependence of the ellipsometric phase of the reference solution ($n=1.33$), and the analyte refractive index $n = 1.33 + \delta n$, $\delta n = 0.001$, calculated at an incidence angle slightly away from the phase singularity $\theta = \theta_{PS} + \delta \theta$, $\delta \theta = -0.05^{\circ}$. 
The shaded area shows the difference $\Delta - \Delta_0$; the black line corresponds to the maximal difference of $|\Delta - \Delta_0|$.
(d) Maximum shift of the ellipsometric phase $\DM$ as a function of the analyte refractive index shift $\delta n$. 
Inset: sensitivity $S = \partial \DM / \partial \delta n$ as a function of $\delta n$.
(e) The influence of noise on the optical phase near the PS according to the model, Eq.~\eqref{Eq_5b}.
}
\label{fig1}
\end{figure}

We begin by outlining of the basic principles underlying the topological phase singularity-based sensing. 
Essentially, the refractometric sensor is represented by a linear reflecting structure covered with an optically thick water solution of the sensing medium (analyte), Figure~\ref{fig1}(a).
A change in the concentration of the sensing medium induces a variation of the analyte refractive index, which in turn causes a change of the reflection intensity that can be detected.
The idea of the ellipsometry-assisted sensing, in contrast to the common surface plasmon-polariton schemes, is to monitor the changes in the ratio of reflected $s$- and $p$-polarized waves $\rho = r_p/r_s$, which can be routinely assessed with spectroscopic ellipsometry \cite{ermolaev2022topological, thomas2022all, sreekanth2018biosensing}.

In our study we examine a system consisting of an anisotropic absorbing substrate and an optically thick water solution of the sensing medium with  refractive index $n$, placed on it, Figure~\ref{fig1}(a).
Optical anisotropy of the substrate will play the crucial role in the behavior of the ellipsometric response $\rho$.
As a model of the anisotropic substrate we consider a uniaxial absorbing crystal with the optical axis perpendicular to the interface. The in-plane permittivity $\eps_{xx} = \eps_{yy} \equiv \eps_{\bot}$ is described by the Lorentz model: 
\begin{equation}
    \eps_\mathrm{Lor} (\w) = \eps_{\infty} + f \frac{\w_0^2}{\w_0^2 - \w^2 - i \gamma \w},
\end{equation}
where $\eps_{\infty}$ is the high-frequency permittivity, $f$ is the oscillator strength  of the resonant transition of the medium, $\w_0$ is its resonant frequency, and $\gamma$ describes its non-radiative decay rate. 
The permittivity along the optical axis is a complex-valued constant, $\eps_{zz} \equiv \eps_{\parallel} = \mathrm{const}$. 

The substrate-analyte interface is illuminated by a linearly polarized plane wave incident at an angle $\theta$, Figure \ref{fig1}(a).
Since in real experiments the thickness of the analyte substantially exceeds the coherence length of optical radiation, reflection from the analyte-substrate interface can be considered as reflection at the interface of two semi-infinite media and described by Fresnel formulas for $s$- and $p$- polarized wave:
\begin{equation}
    \begin{split}
        & r_s = \frac{ k_{z} - k_{z}^{(o)}}{ k_{z} + k_{z}^{(o)}}, \\
        & r_p = \frac{ k_{z}/n^2 - k_{z}^{(e)}/\eps_\perp(\w) }{ k_{z}/n^2 + k_{z}^{(e)}/\eps_\perp(\w) }, 
    \end{split}  
\end{equation}
where $k_{z} = \omega/c \sqrt{n^2-\sin^2\theta}$ is the $z$-component of the wave vector in the analyte, $k_{z}^{(o)} = \w/c \sqrt{\eps_\perp(\w)-\sin^2\theta}$ and $k_{z}^{(e)} = \omega/c \sqrt{\eps_\perp(\omega)/\eps_\parallel (\eps_\parallel - \sin^2\theta)}$ are the $z$-components of the wave vectors of the $s$- and $p$-polarized (ordinary, $(o)$, and extraordinary, $(e)$) transmitted waves in the uniaxial medium.
Reflection of the incident wave at the air-analyte interface slightly reduces the intensity of the transmitted wave, but barely affects the ellipsometry measurements, as demonstrated in a previous work \cite{ermolaev2022topological}.

As shown previously, a semi-infinite absorbing uniaxial crystal may exhibit points of zero reflection for $p$-polarized excitation \cite{Baranov2015c, maslova2024topological} (see Supplementary Material, Fig. S1).
Owing to the point of zero reflection, the ellipsometric response $\rho$ exhibits a phase singularity at a point $\omega_{PS}, \theta_{PS}$ of the frequency-incidence angle space $\omega$, $\theta$, where the ellipsometric phase $\Delta = \arg \rho$ becomes undetermined, Figure~\ref{fig1}(b).
At a fixed incidence angle close to $\theta_{PS}$ the argument of complex-valued $\rho$ demonstrates a rapid variation in a narrow frequency range in the vicinity of the singularity.  
This implies that a weak change of the optical environment (e.g., the analyte index) may lead to a significant change of the response function argument, Figure~\ref{fig1}(c). 
The core idea of the topological singularity-based sensor is to take advantage of this rapid variation of the ellipsometric phase near the point of zero reflection of $p$-polarized wave.


In order to detect a change of the analyte refractive index, one 
(i) measures the ellipsometric phase $\Delta(\omega)$ near the phase jump,
and (ii) calculates the difference between the measured values and the initial calibration curve $\Delta_0(\omega)$ corresponding to the analyte refractive index, Figure~\ref{fig1}(c).
Next, we create the correspondence between the known refractive index shift of the analyte $\delta n$ and the maximum absolute deviation of the ellipsometric phase $\max ( |\Delta(\w) - \Delta_0 (\w)| )$ over a range of frequencies:
\begin{equation}
    \DM = \max_{\w_{\min} < \w < \w_{\max}}
    ( |\Delta(\w) - \Delta_0 (\w)| ),
\end{equation}
where the range $\w_{\min}$ to $\w_{\max}$ is determined by the experimental limitations. For this sudy we chose $\w_{\min} = 0.49 \,\omega_0$ and $\w_{\max} = 0.52 \, \omega_0$ ($\omega_{PS} \approx 0.505 \,\omega_0$).
This correspondence allows us to determine $\delta n$ given an experimentally measured value of $\DM$, as Figure~\ref{fig1}(d) demonstrates.
Tangent of the curve locally determines the sensitivity $S$ of the system to the changes of the analyte index:
\begin{equation}
    S = \frac{\partial \DM }{\partial \delta n}.
    \label{Eq_5}
\end{equation}

One could naively incorporate noises in this model by adding a randomly distributed noise term to the real and imaginary parts of s- and p-polarized reflection amplitudes:
\begin{equation}
    \Delta' = \arg \frac{r_p + \delta_1 + i \delta_2}{r_s + \delta_3 + i \delta_4},
    \label{Eq_5a}
\end{equation}
where $\delta_i$ is a randomly distributed noise-induced quantity added to the real and imaginary parts of the corresponding amplitude.
Close to zero of $r_p$ and as long as $r_s$ does not approach zero at the same time, we can approximate
\begin{equation}
    \Delta'   \approx     \arg \frac{r_p + \delta_1 + i \delta_2}{r_s} \to \arg (\delta_1 + i \delta_2) - \arg {r_s}.
    \label{Eq_5b}
\end{equation}
As expected, the argument of the remaining noise term produces a highly distorted magnitude of $\Delta$ close to the phase singularity, Figure~\ref{fig1}(e). 
In reality, however, incorporating noise sources into the ellipsometric scheme is a more non-trivial problem compared to the naive approach.


\subsection*{Phase measurement in ellipsometry}

\begin{figure}[t!]
\centering\includegraphics[width=1.0\textwidth]{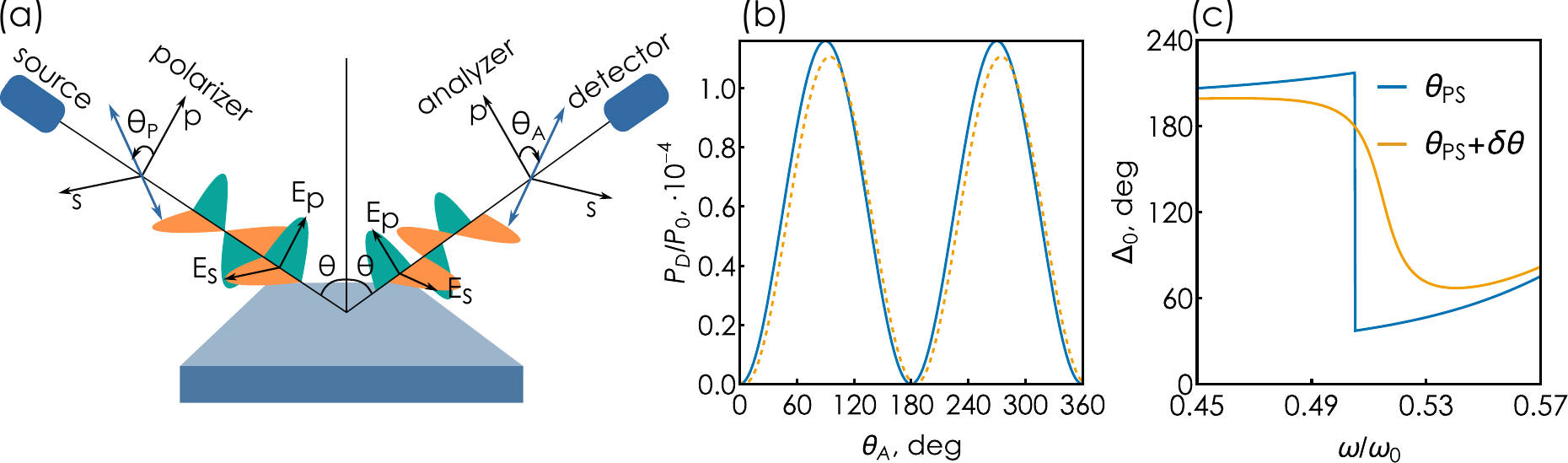}
\caption{\textbf{Ellipsometry basics.} 
(a) The principal scheme of spectroscopic ellipsometry measurement process. An unpolarized quasi-monochromatic light from the source passes through a polarizer, reflects off of the interrogated substrate, passes through an analyzer and gets recorded at the detector.
(b) Recorded power $P_D$ at the phase singularity frequency $\omega=\omega_{PS}$ for different incidence angles: $\theta=\theta_{PS}$ and $\theta=\theta_{PS} + \delta \theta$, where $\delta \theta$ is set to be $\delta \theta = -2^{\circ}$. 
(c) The ellipsometric phase behavior near $\omega=\omega_{PS}$ for the same incidence angles as in (b).}
\label{fig2}
\end{figure}

To offer a more accurate picture of the noise mechanism arising in an sensing experiment, we first briefly discuss the key steps of ellipsometric spectroscopy, schematically illustrated in Figure~\ref{fig2}(a). An unpolarized light from a lamp source passes through a monochromator and next through a polarizer set at an angle $\theta_P$, thus becoming linearly polarized and acquiring both $s$ and $p$ components with respect to the substrate.
In the following calculations we set $\theta_P$ to $\pi/4$, such that transmitted light has equally weighted $s$ and $p$-polarized components. 
After reflection from the substrate, the wave passes through an analyzer with a rotating axis characterized by the angle $\theta_A$.
Power of the signal reaching the detector reads:
\begin{equation}
     P_D = \frac{P_0 (|r_p|^2 \cos^2\theta_P + |r_s|^2 \sin^2\theta_P)}{4} \left(1 + \alpha \cos{2\theta_A} + \beta \sin{2\theta_A}\right), 
\label{Eq_7}
\end{equation}
where $P_0$ is the power of the lamp source, and $\alpha$ and $\beta$ are expressed via the desired ellipsometric parameters:
\begin{equation}
    \begin{split}
        & \alpha = \frac{\tan^2\Psi-\tan^2\theta_P}{\tan^2\Psi+\tan^2\theta_P} \\
        & \beta = \frac{2\tan{\Psi}\cos{\Delta}\tan \theta_P}{\tan^2\Psi+\tan^2\theta_P},
    \end{split}
    \label{Eq_9}
\end{equation}
with $\rho = r_p/r_s \equiv \tan{\Psi}\exp{(i\Delta)}$.


The detector records the signal for different analyzer angles. 
Figure~\ref{fig2}(b) shows an example of detected signal as a function of the analyzer angle. Applying the Fourier transform to this data allows one to determine the coefficients $\alpha$ and $\beta$ in Eq. 6, which in turns allows us to calculate the desired ellipsometric magnitude and phase:
\begin{equation}
    \tan \Psi = \sqrt{\frac{1+\alpha}{1-\alpha}}|\tan{\theta_P}| 
\end{equation}
\begin{equation}
    \cos{\Delta} = \frac{\beta}{\sqrt{1-\alpha^2} }\frac{\tan{\theta_P}}{|\tan{\theta_P}|}
\end{equation}

The choice of the incidence angle profoundly affects the initial calibration curve $\Delta_0 (\w)$.
As can be seen from Figure~\ref{fig2}(b) and Figure~\ref{fig2}(c), 
a deviation of the incidence angle from the exact phase singularity $\theta_{PS}$ only slightly alters the power $P_D$, but that leads to a noticeable change of the calibration curve.
As we will see in the following, the choice of the calibration curve profoundly affects the trade-off between the sensitivity and the resolution of the ellipsometry-assisted refractometric sensor.


\subsection*{Noise model}

\begin{figure}[t!]
\centering\includegraphics[width=1.0\textwidth]{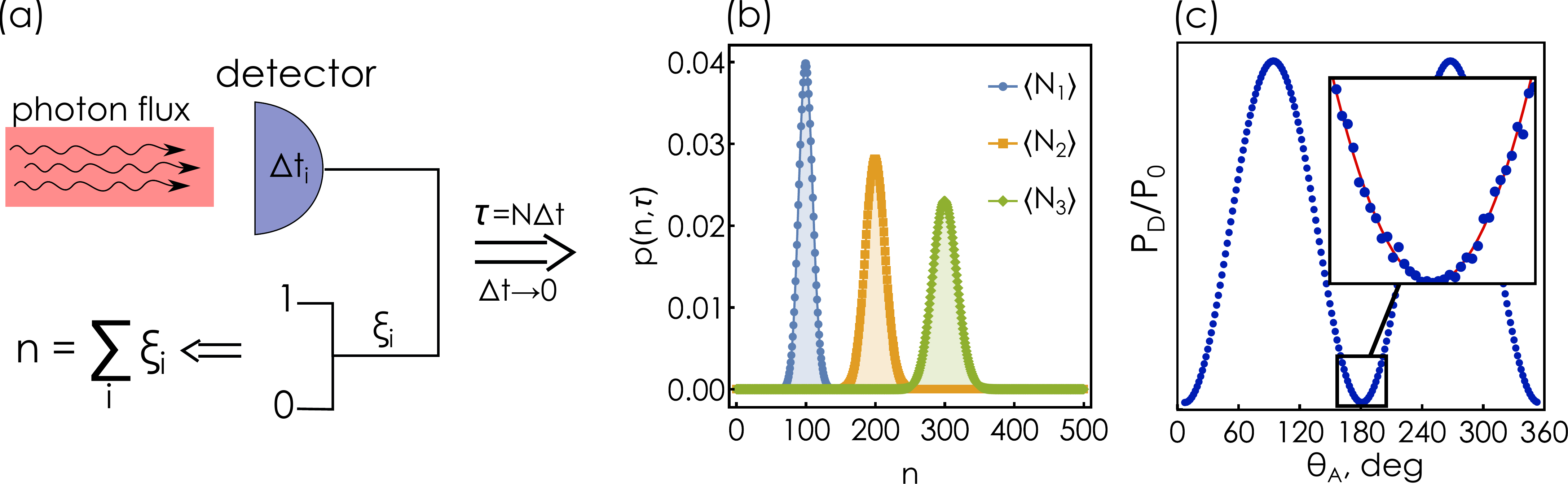}
\caption{\textbf{Shot noise model.} 
(a) Probabilistic registration of photoelectrons: incoming photon flux in each time interval $\delta \tau$ can excite a photoelectron with probability $p_0$, which results in $n$ detected photoelectrons during the entire measurement interval.
(b) Probability density of recording $n$ photoelectrons over a finite period time $\tau$ having a Poisson distribution for different expected values $\langle N_1 \rangle < \langle N_2 \rangle < \langle N_3 \rangle$. 
(c) Signal recorded by the detector as a function of the analyzer angle $\theta_A$ in the presence of shot noise.}
\label{fig3}
\end{figure}

Next we develop an analytical model of noise that can be adopted for modeling the refractometric sensing process, illustrated in Figure~\ref{fig3}(a).
Assume the detector is subject to an incoming photon stream reflected from the substrate with constant power $P_D$ given by Eq.~\eqref{Eq_7}.
Since photoelectron excitation is a probabilistic process, the number of recorded counts varies. 
This variation of the detected photon counts is the manifestation of shot noise \cite{MandelLeonardandWolf1995}.
The process of counting photoelectrons is described by the probability density $p(n,\tau)$ that describes the probability of recording $n$ photoelectrons during the measurement time interval $\tau$.

To determine the probability density $p(n,\tau)$, following \cite{carmichael2009open} we divide the measurement interval $\tau$ into $M$ infinitesimal intervals $\delta t = \tau/M$ , such that in each interval it is possible to register at most one photoelectron with elementary probability $p_0 = \eta P_D \delta t/(\hbar \omega)$ where $\hbar \w$ is the photon energy and $\eta$ is the detector quantum efficiency. 
Probability of recording $n$ photoelectrons during the whole measurement interval $\tau$ is determined by the statistics of $M$ independent events and is described by the binomial distribution:
\begin{equation}
\begin{split}
    &p(n,\tau) = \frac{M!}{n!(M - n)!} p_0^n (1 - p_0)^{M - n} = \\
    &\frac{M(M-1)...(M - n + 1)}{M^n} \frac{(\eta P_D M \delta t/(\hbar \omega))^n}
    {n!} \times \\
    &\times (1 - (\eta P_D \delta t/(\hbar \omega))) ^ {M - n}.
\end{split}
\end{equation}
In the limit $\delta t \rightarrow 0$, $M \rightarrow \infty$ this approaches the Poisson distribution (Figure~\ref{fig3}(b)):
\begin{equation}
    p(n,\tau) = \frac{\langle N \rangle ^n}{n!} \exp (- \langle N \rangle),
\end{equation}
where the expected number $\langle N \rangle$ of registered photon counts reads:
\begin{equation}
    \langle N \rangle = \eta \frac{P_D}{\hbar \omega} \tau,
    \label{Eq_6}
\end{equation}

The probabilistic distribution of detected photoelectrons leads to the variation of the signal recorded by the detector as a function of the analyzer angle, Figure~\ref{fig3}(c).
The relative uncertainty of these variations is particularly noticeable around $\theta_A = \pi n$, where the analyzer passes only (vanishingly small due to phase singularity) $p$-polarized component of the reflected field. This large relative uncertainty $\propto 1/ \sqrt{\langle N \rangle}$ is a natural property of a Poisson-distributed variable with low expected value.
However, this uncertainty rapidly vanishes as the incidence angle gets detuned from the angle of phase singularity.

This noise in the recorded number of photoelectrons leads to uncertainties in determining the Fourier coefficients $\alpha$ and $\beta$, Eq. \eqref{Eq_9}, which in turn determine the magnitude $\tan \Psi$ and the sought for ellipsometric phase $\Delta$.
However, as we will see in the following it is performing the ensemble of measurements for different analyzer angles that enables stable evaluation of the analyte index despite the seeming issue of absent photon flux at the phase singularity.

\subsection*{The effect of noise in the experiment simulation}

To render the combined numerical-theoretical simulations more realistic and to grasp the scale of the physical parameters where the influence of noise becomes significant, we will set the numerical values of the quantities characterizing the actual experiment.
A lamp with a monochromator is used as a quasi-monochromatic light source. In simulations, we consider the typical range of spectral power densities $P$ ranging from 1~$\rm{\mu W/nm}$ to 100~$\rm{\mu W/nm}$ at the wavelength of the phase singularity $\lambda_{PS} = 600$ nm. 
The monochromator filters out a narrow wavelength band from the entire spectrum of bandwidth $\Delta \lambda = 5~\text{nm}$ centered at a given wavelength.



\begin{figure}[t!]
\centering\includegraphics[width=1.0\textwidth]{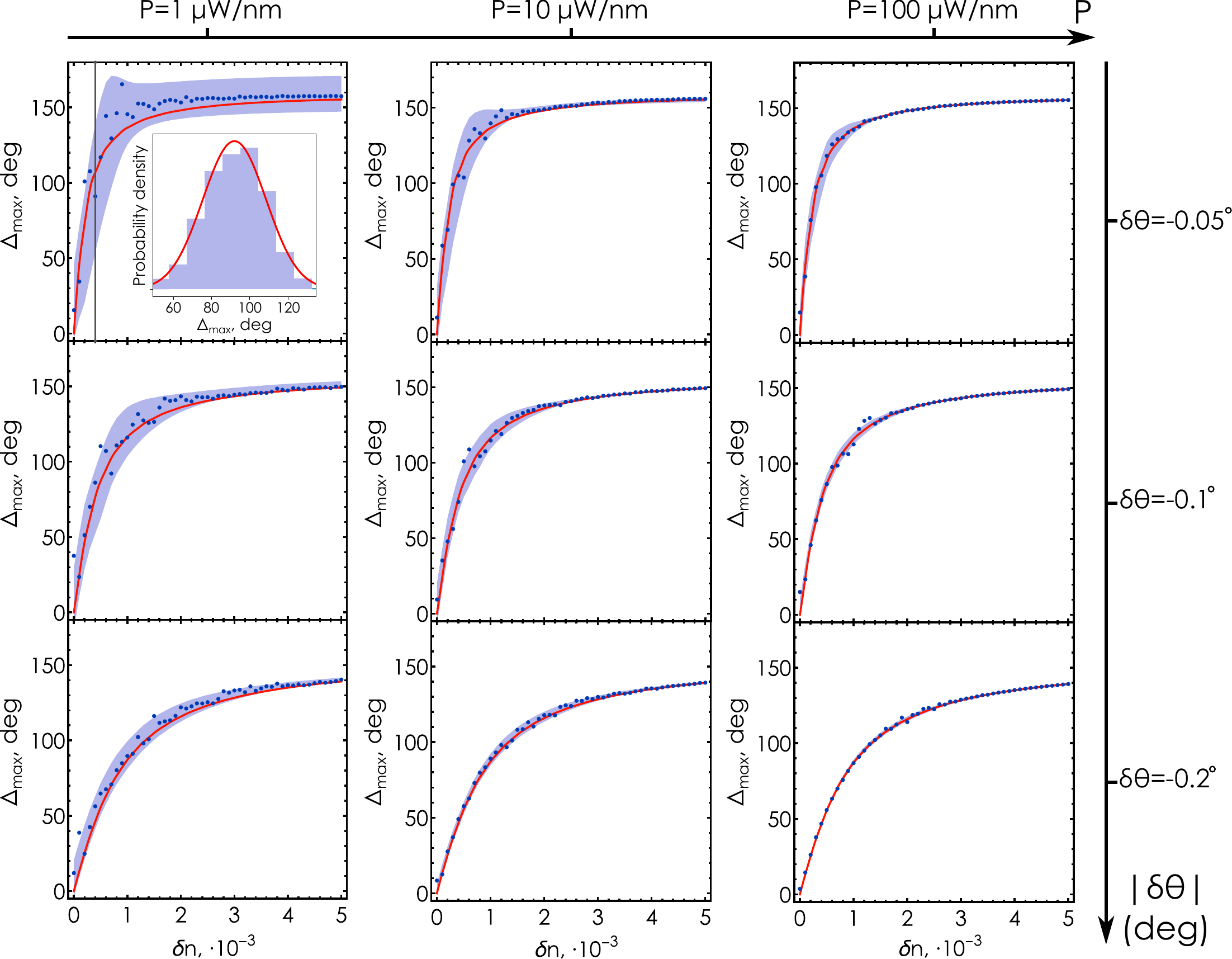}
\caption{\textbf{The effect of shot noise on refractometric sensing.} 
Behavior of  noisy sensitivity with increasing spectral power $P$ (from left to right) and increasing deviation $|\delta \theta|$ of the incidence angle $\theta = \theta_{PS}+\delta \theta$ from the angle of phase singularity $\theta_{PS}$ (from top to bottom). 
Each dot represents the result of a single simulated experiment. 
The gray area represents the corridor of values into which each point falls with a probability of 95\%. 
The red curve shows the ideal phase shift $\DM$ obtained in the noise-free model.Inset: example of distribution of $\DM$ for $\delta n = 4\cdot 10^{-4}$ (vertical gray line) obtained from a numerical simulation of the experiment for $P = 1\,\rm{\mu W/nm}$ and $|\delta \theta| = 0.05^\circ$ in comparison with the normal distribution (red curve).
}
\label{fig4}
\end{figure}

Now we are in a position to analyze the influence of the spectral power and the incidence angle on the performance of the ellipsometry-based refractometric sensing close to the phase singularity point.
To that end we run a set of numerical "experiments" characterized by different spectral powers and incidence angles and fixed measurement interval $\tau = 0.2$ s, Figure~\ref{fig4}. 
For each combination of incident power and incidence angle we run the simulation repeatedly many times (in our calculations, the number of repetitions is 500). 
For each $\delta n$, we obtain a set of values of the maximum ellipsometric phase shift $\DM$, for which we can calculate the mean $\langle \DM \rangle$ and standard deviation $\sigma_\Delta$ (see Supplementary Material, Figs. S2 and S3). 
Notably, the mean measured value $\langle \DM \rangle$ deviates substantially from the etalon phase shift $\DM^{(0)}$ obtained in the noise-free model, $\langle \DM \rangle \ne \DM^{(0)}$, see Supplementary Material, Fig. S2.

As we find from our numerical simulations, the probability distribution of $\DM$ for each $\delta n$ follows the normal distribution with a reasonable accuracy (see Supplementary Material, Figs. S4 -- S9).
The mean values and standard deviations can be approximated by smooth curves that are functions of the analyte medium refractive index shift $\delta n$ and characterize the impact of shot noise on the measurement. 
This approximation allows us to visualize for each $\delta n$ an interval $\langle \DM \rangle \pm \widetilde{\sigma}_\Delta$, where a simulated "measurement" falls with a probability of 95\% (shaded regions in Figure~\ref{fig4}) with $\widetilde{\sigma}_\Delta = \sqrt{2} \sigma_\Delta \text{erf}^{-1}(0.95)$ and $\text{erf}^{-1}(x)$ being the inverse of the error function. Figure \ref{fig4} clearly demonstrates that this distribution interval characterizing the influence of shot noise can be reduced either by increasing the source spectral density $P$, which is limited in the experiment, or by shifting the incidence angle away from the phase singularity value $\theta_{PS}$.

\begin{figure}[t!]
\centering\includegraphics[width=1.0\textwidth]{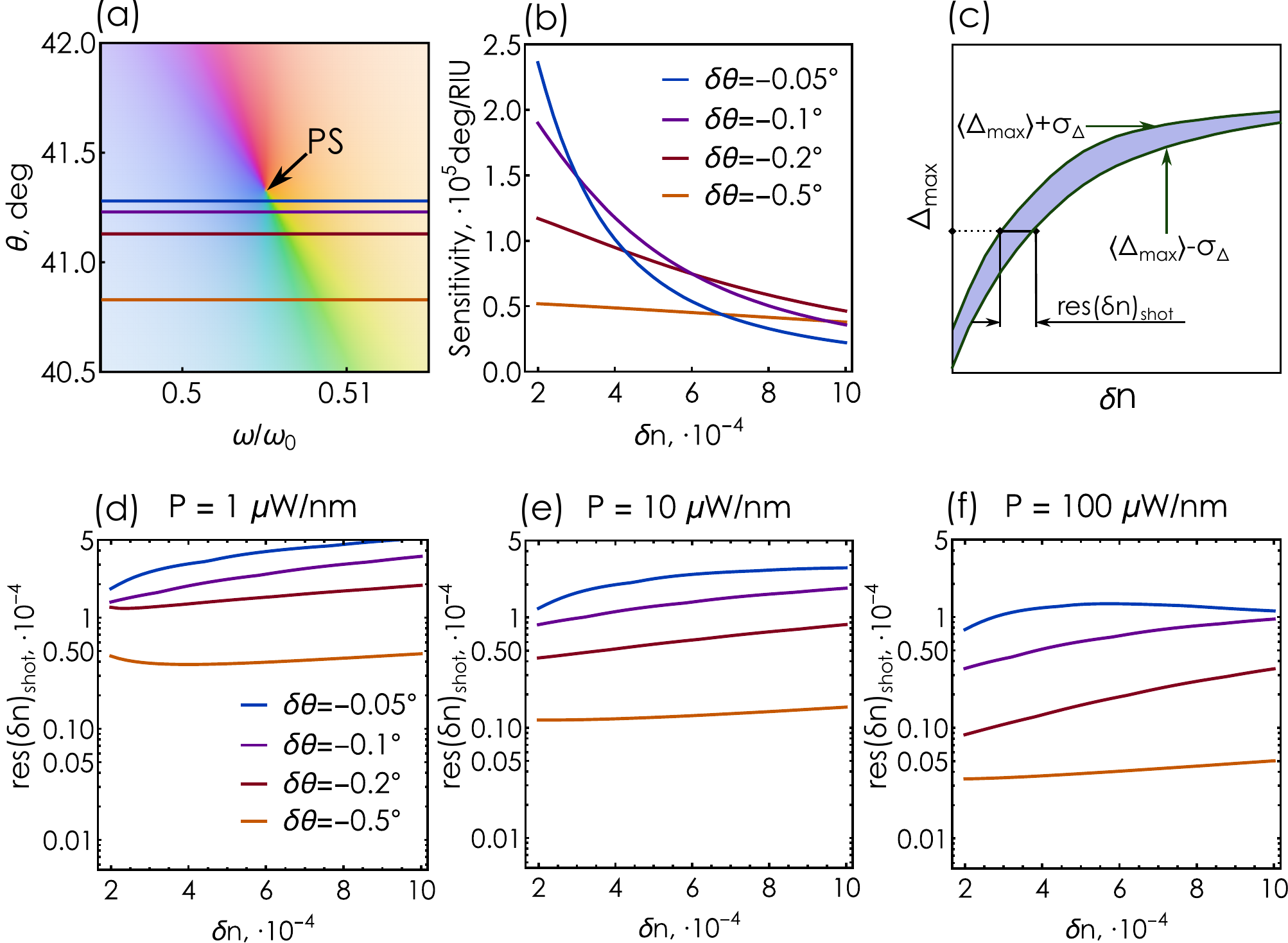}
\caption{\textbf{The trade-off of sensitivity and resolution in the ellipsometry-based sensor.} (a) Incidence angles $\theta = \theta_{PS} + \delta\theta$ characterized by different deviations $|\delta\theta|$ from the angle $\theta_{PS}$ and defining calibration curves. (b) The sensitivity of the sensor as a function of $\delta n$ for deviation angles corresponding to the incidence angles shown in (a). (c) Graphical definition of the noise-induced resolution $\mathrm{res}(\delta n)_\mathrm{shot}$. (d) The resolution of $\delta n$ for the incidence angles shown in (a) for a source spectral power $P = 1\, \rm{\mu W/nm}$. (e) Same as (d) for $P = 10 \,\rm{\mu W/nm}$. (f) Same as (d) for $P = 100\, \rm{\mu W/nm}$.}
\label{fig5}
\end{figure}

This observation allows one to use a phase singularity-based sensor even with a low-power optical source. However, deviating the incidence angle away from the singularity to achieve the required accuracy comes at the cost of reduced \emph{sensitivity}.
Figure~\ref{fig5}(b) shows the behavior of the sensitivity $S$ on $\delta n$ for a few different values of the incidence angles close to the phase singularity.
Deviating the incidence angle away from $\theta_{PS}$ results in a smoother sensitivity behavior. 
At the same time, in the region of weak variation of the analyte  refractive index ($\delta n < 3\cdot 10^{-4}$), the sensitivity drops from $2\cdot 10^5$ deg/RIU to $0.5 \cdot 10^5$ deg/RIU upon a slight $0.5^{\circ}$ deviation of the incidence angle.

Next, we study the effect of the incidence angle and the source power on the resolution of the refractometric sensor.
We define the shot noise-limited resolution $\mathrm{res}(\delta n)_\mathrm{shot}$ for a given measured value $\langle \DM \rangle$ as the width of the region bounded by the curves $\langle \DM \rangle \pm \sigma_\Delta$, Figure~\ref{fig5}(c). 
Figures \ref{fig5}(d-f) show the behavior of the noise-limited resolution at different incidence angles and source spectral powers.
Expectedly, increasing the source power improves the noise-limited resolution for any incidence angle.
Increasing the deviation angle for a fixed power improves the resolution from $\sim 2\cdot 10^{-4}$ to $\sim 5\cdot 10^{-5}$ even at extremely low source powers of $P = 1\,\rm{\mu W/nm}$. 
In addition, improving the resolution by increasing $|\delta \theta|$ (and not the power) can be more effective, as shown by the curves corresponding to $|\delta \theta| = 0.05^{\circ}$; 
with an increase in power by two orders of magnitude, the resolution remains relatively crude ($\sim 10^{-4}$).
As a result, one can substantially increase the sensor robustness by a slight change of the incidence angle without upgrading the light source to high-power version.

As a final step of our analysis, we note that sensitivity imposes another potential constraint on the resolution of our system even in the absence of shot noises. As a rule of thumb, an ellipsometer can reliably detect an ellipsometric phase shift $\DM \gtrsim 1^\circ$ \cite{ermolaev2022topological}. This imposes another constraint on the resolution that can be easily estimated as
\begin{equation}
    \mathrm{res} (\delta n)_\mathrm{sens} = \frac{1^\circ}{S}
\label{Eq_15}
\end{equation}
with $S$ being the sensitivity of the setup, Eq. \eqref{Eq_5}.
As shown above, while increasing $|\delta \theta|$ improves the noise-limited resolution $\mathrm{res} (\delta n) _\mathrm{shot}$, it at the same time reduces the sensitivity. 
This causes the sensitivity, not the shot noise, to be the decisive factor determining the resolution at relatively large $|\delta \theta|$.
For our model system an increase of $|\delta \theta|$ up to $0.5^{\circ}$ significantly reduces the sensitivity and becomes the decisive resolution factor at source powers exceeding $P = 10 \,\rm{\mu W/nm}$ (see Supplementary Material, Fig. S10).
Nevertheless, even then the sensitivity is sufficient to detect a refractive index variation of $\approx 10^{-4}$ accompanied by the ellipsometric phase shift $\Delta_{\max}$ of more than $5^{\circ}$.

\section{Discussion and Conclusion}

To conclude, we have developed an accurate model of an ellipsometry-based refractometric sensor operating close to a topological phase singularity at the point of zero reflection. 
The developed model accounts for the shot noise caused by the discrete nature of the detected photon stream. Our results suggest that such an ellipsometry-based refractometric sensor remains robust even in the presence of shot noise near the point of zero reflection.
This approach allows one to use a phase singularity-based refractometric sensor even at relatively low source powers, making the system robust against noise and offering a route toward their implementation in wearable compact devices.

The results reveal that the incidence angle provides a convenient control knob to manipulate the sensitivity and resolution of the sensing process.
However, it is crucial to note that deviation from the phase singularity reduces the sensitivity of the system, which in turn inhibits the resolution.
This introduces a complex balance between the sensitivity and the resolution of the ellipsometric sensor.
When considering an actual system with specified characteristics, the developed model will allow choosing the optimal value of the incidence angle and determining the minimum possible recorded value of refractive index change with the desired accuracy.

\section{Acknowledgments}

The work was supported by the Ministry of Science and Higher Education of the Russian Federation (FSMG-2024-0014).
D.G.B. and V.M. acknowledge support from Russian Science Foundation (grant No. 23-72-10005). D.G.B. acknowledges support from BASIS Foundation (grant No. 22-1-3-2-1). V.M. acknowledges support from BASIS Foundation (grant No. 24-1-5-136-1). E.S.A. acknowledges the support of the Foundation for the Advancement of Theoretical Physics and Mathematics BASIS.


\bibliography{sensor}

\end{document}